\documentclass[a4paper,fleqn,usenatbib]{mnras}
\usepackage[T1]{fontenc}
\usepackage{ae,aecompl}
\usepackage{comment}

\usepackage{microtype}
\usepackage{amsmath}
\usepackage{graphicx}
\usepackage{lscape}

\usepackage{amssymb}
\usepackage{color}
\usepackage[flushleft]{threeparttable}

\newcommand {\bc}{\begin {center}}
\newcommand {\ec}{\end {center}}
\newcommand {\be}{\begin {equation}}
\newcommand {\ee}{\end {equation}}
\newcommand {\beq}{\begin {eqnarray}}
\newcommand {\eeq}{\end {eqnarray}}

\def\flux{erg s$^{-1}$ cm$^{-2}$}
\def\lum{erg s$^{-1}$}

\def\gx{GX\,304$-$1}
\def\a05{A\,0535$+$262}

\title[Very low state of \a05]
{Cyclotron emission, absorption, and the two faces of X-ray pulsar \a05}

\author[S.~S.~Tsygankov et al.]
{Sergey~S.~Tsygankov,$^{1,2}$\thanks{E-mail: sergey.tsygankov@utu.fi} 
Victor Doroshenko,$^{3,2}$
Alexander A.~Mushtukov,$^{4,2,5}$\newauthor 
Valery F. Suleimanov,$^{3,6,2}$
Alexander A. Lutovinov,$^{2,7}$
and Juri~Poutanen$^{1,2}$
 \\
$^1$Department of Physics and Astronomy,  FI-20014 University of Turku, Finland \\
$^2$Space Research Institute of the Russian Academy of Sciences, Profsoyuznaya Str. 84/32, Moscow 117997, Russia \\
$^3$Institut f\"ur Astronomie und Astrophysik, Universit\"at T\"ubingen, Sand 1, D-72076 T\"ubingen, Germany \\
$^4$Leiden Observatory, Leiden University, NL-2300RA Leiden, The Netherlands\\
$^5$Pulkovo Observatory, Russian Academy of Sciences, Saint Petersburg 196140, Russia\\
$^6$Kazan (Volga region) Federal University, Kremlevskaya str. 18, 420008 Kazan, Russia\\
$^7$Higher School of Economics, Myasnitskaya 20, 101000 Moscow, Russia }

\date{Accepted 2019 May 22. Received 2019 May 22; in original form 2019 April 3}

\pubyear{2019}

\begin{document}
\label{firstpage}
\pagerange{\pageref{firstpage}--\pageref{lastpage}}
\maketitle

\begin{abstract}
Deep {\it NuSTAR} observation of X-ray pulsar \a05, performed at a very low luminosity of $\sim7\times10^{34}$~\lum, revealed the presence of two spectral components. 
We argue that  the  high-energy component is associated with cyclotron emission from recombination of electrons collisionally excited to the upper Landau levels.
The cyclotron line energy of $E_{\rm cyc}=47.7\pm0.8$~keV was measured at the luminosity of almost an order of magnitude lower than what was achieved before. The data firmly exclude a positive correlation of the cyclotron energy with the mass accretion rate in this source.
 
\end{abstract}

\begin{keywords}
{accretion, accretion discs -- pulsars: general -- scattering --  stars: magnetic field -- stars: neutron -- X-rays: binaries }
\end{keywords}

\section{Introduction}
\label{intro}

Physical processes dominating the X-ray emission from accreting neutron stars
(NSs) depend strongly on the properties of the NS and the mass accretion rate.
X-ray pulsars (XRPs), constituting a separate subclass of accreting NSs, are
especially interesting in this respect allowing us to study microphysics under
extreme conditions of ultrastrong magnetic fields. Observations of such systems
at very low luminosities are of particular interest as they 
reveal the emission region in its  simplest form of the hotspot at the NS surface, allowing us to avoid modeling of several
complex effects associated with potential formation of an accretion column, propagation of light from the column to the observer, and 
illumination of the NS surface.
 However, even in this case the nature brought some surprises. In particular,
such observations revealed dramatic transition from a cut-off power-law
spectrum typical for XRPs at high luminosities to a two-component spectrum with
peaks around 5 and 40\,keV at an accretion rate as low as $\sim10^{14}$ g
s$^{-1}$ in a transient pulsar \gx\ \citep{2019MNRAS.483L.144T}. Similarly to
another low-luminosity XRP, X Persei, such spectral shape cannot be explained
with the existing models developed for high accretion rate sources \citep[see
e.g.][]{2007ApJ...654..435B, 2016A&A...591A..29F}.

At very low rates the emission is produced at the NS surface and allow  to study the interaction of the infalling material with the highly magnetized atmosphere. 
Several  mechanisms are able to explain the existence of the high-energy spectral component, with the most plausible being (i) the cyclotron emission reprocessed by the magnetic Compton scattering or (ii) the thermal radiation of deep atmospheric layers partly Comptonized in the overheated upper layers. 
In addition to \gx\ and  X Persei, more examples of low-luminosity states need to be observed before the final conclusion on the physical mechanisms behind spectrum formation can be made.

In this Letter we consider one of the best studied Be/X-ray binaries, \a05, ideally suitable for such kind of studies. 
The source contains a slowly rotating NS \citep[$P_{\rm spin}\approx104$~s; ][]{1975Natur.256..628R} orbiting around B0IIIe star HDE~245770 at distance around 2 kpc from the Sun \citep{1998MNRAS.297L...5S}. 
A relatively long orbital period $P_{\rm orb}\approx110$~d \citep{1996ApJ...459..288F} implies that the source stays in the quiescence for a long period of time between periastron passages. 
Previously the source was observed in deep quiescence state with luminosity around or below $\sim10^{34}$~\lum\ several times with different instruments \citep{2000A&A...356.1003N,2004NuPhS.132..476O,2013ApJ...770...19R,2014A&A...561A..96D}. It was demonstrated  \citep[see e.g.][]{2014A&A...561A..96D}  that the emission even at  such a low luminosity is dominated by the accretion from the ``cold'' accretion disc \citep{2017A&A...608A..17T}.
However, the low sensitivity  of  the available data above 10 keV did not allow to study the broad band spectrum of \a05\ at this luminosity. 
For this purpose we performed deep {\it NuSTAR} observations of \a05\  between two type-I outbursts when the luminosity was as low as $\sim7\times10^{34}$ \lum. 

To determine the physical origin of the spectral components it is crucial to know the strength of the NS magnetic field. In the case of \a05\ it is known from the cyclotron resonant scattering feature (CRSF) energy of $\sim45$ keV and its first harmonic around 100 keV \citep{1994A&A...291L..31K,1995ApJ...438L..25G,1996A&AS..120C.175K}. Unlike some others low-luminosity XRPs, \a05\ exhibits no significant positive correlation of the cyclotron line energy with luminosity in the pulse-averaged spectra (\citealt{2007A&A...465L..21C,2013ApJ...764L..23C,2017A&A...608A.105B}, however, see \citealt{2015ApJ...806..193S}). At the same time such correlation was observed in the pulse-amplitude-resolved spectroscopy \citep{2011A&A...532A.126K}. 
Investigation of the cyclotron energy behaviour at very low luminosity is another topic we address below.

\section{Data analysis}
\label{sec:data}

In this work we use three \textit{NuSTAR} \citep{2013ApJ...770..103H} observations of the source performed in wide range of luminosities. Two brighter observations were taken in the declining phase of the outburst in the beginning of 2015 \citep{2017A&A...608A.105B}, and the third one -- on 2018 December 26 (ObsID 90401370001), in a deep quiescent state, $\sim$3 months after the outburst that ended in October 2018. 

The raw data were processed following the standard data reduction
procedures described in the {\it NuSTAR} user guide, and using the standard {\it NuSTAR} Data Analysis Software {\sc nustardas} v1.8.0 provided under {\sc heasoft v6.25} with the CALDB version 20181022.  
The source and background spectra were extracted from the circular regions with radii of 50\arcsec\ and 90\arcsec, respectively, using the {\sc nuproducts} routine. The background was extracted from a source-free region in the corner of the field of view. 
Final spectra were optimally rebinned using the prescription in \cite{2016A&A...587A.151K}.

\begin{table}
        \caption{Observational log of \a05. }
        \label{tab:log}
        \centering
        \begin{tabular}{lccc}
                \hline
                ObsID & $T_{\rm start}$ & $T_{\rm stop}$ & Exposure \\
                      &  MJD         &   MJD       &  ks      \\
                \hline
\multicolumn{4}{c}{{\it NuSTAR} observations}\\
                80001016002 & 57064.17 & 57064.68 & 21.5 \\
                80001016004 & 57067.07 & 57067.73 & 29.7 \\
                90401370001 & 58478.11 & 58479.51 & 54.9 \\
                \hline
\multicolumn{4}{c}{{\it Swift} observations}\\
                00081432001 & 57064.17 & 57064.24 & 1.9 \\
                00035066052 & 57068.89 & 57068.90 & 1.1 \\
                00088834001 & 58479.60 & 58479.67 & 1.9 \\
                \hline
        \end{tabular}
\end{table}

To expand our spectral analysis to the softer energy band, we also used  the data from the XRT telescope \citep{2005SSRv..120..165B} on-board the {\it Neil Gehrels Swift Observatory} \citep{2004ApJ...611.1005G} obtained simultaneously or very close in time to the {\it NuSTAR} observations. 
The observational log can be found in Table~\ref{tab:log}.

The spectral extraction from the XRT data was done using the online tools
\citep[][]{2009MNRAS.397.1177E}\footnote{\url{http://www.swift.ac.uk/user_objects/}}
provided by the UK Swift Science Data Centre. 
To fit the spectra in the {\sc xspec} package we binned the spectra to have at least 1 count per energy bin and fitted them using W-statistic
\citep{1979ApJ...230..274W}.
The data from {\it Swift}/XRT and {\it NuSTAR} were used in the 0.3--10 keV and 3.5--79~keV bands, respectively.

\section{Results}

\begin{table*}
        \begin{center}
        \caption{Best-fitting results with {\sc phabs(gau+comptt+comptt)gabs} model  to the broad-band spectra of  \a05\   in different states obtained with {\it NuSTAR} and {\it Swift}/XRT. }
        \label{tab:spe}
        \begin{tabular}{lcccccc}
\hline
  Parameter$^{a}$  & \multicolumn{2}{c}{Low state} & \multicolumn{2}{c}{Medium state} & \multicolumn{2}{c}{High state} \\
\hline
                   & Low-$E$ part & High-$E$ part & Low-$E$ part & High-$E$ part & Low-$E$ part & High-$E$ part\\
$n_{\rm H}$, $10^{22}$ cm$^{-2}$  &  \multicolumn{2}{c}{$0.08\pm0.07$} &  \multicolumn{2}{c}{$0.06\pm0.05$} &  \multicolumn{2}{c}{$0.08\pm0.02$} \\[1ex]
$T_0$, keV                      &  \multicolumn{2}{c}{$0.79\pm0.05$} & \multicolumn{2}{c}{$0.80\pm0.02$} & \multicolumn{2}{c}{$0.74\pm0.01$} \\[1ex]
$T_{\rm p}$, keV$^b$                &  $2.4\pm0.3$   &  $14.5\pm1.1$ &  $2.9\pm0.3$   &  $9.8\pm0.3$ &  $2.4\pm0.4$   &  $10.4\pm0.2$ \\[1ex]
$\tau_{\rm p}$                  &  $6.8\pm0.2$   &  $>10$ &  $7.4\pm0.2$   &  $>10$ &  $>10$   &  $4.5\pm0.6$ \\[1ex]
$E_{\rm cyc}$, keV              &  \multicolumn{2}{c}{$47.7\pm0.8$} &  \multicolumn{2}{c}{$45.8\pm0.4$} &  \multicolumn{2}{c}{$46.1\pm0.2$} \\[1ex]
$W_{\rm cyc}$, keV              &  \multicolumn{2}{c}{$12.6\pm1.5$} &  \multicolumn{2}{c}{$9.6\pm0.5$} &  \multicolumn{2}{c}{$11.3\pm0.3$} \\[1ex]
$\tau_{\rm cyc}$                &  \multicolumn{2}{c}{$2.4\pm0.5$} &  \multicolumn{2}{c}{$1.1\pm0.1$} &  \multicolumn{2}{c}{$0.94\pm0.03$} \\[1ex]
$F_{\rm X}$, $10^{-10}$ \flux   &  \multicolumn{2}{c}{ $1.4\pm0.2$} &  \multicolumn{2}{c}{ $13\pm3$} &  \multicolumn{2}{c}{ $36\pm5$} \\[1ex]
$\chi^2_{\rm red}$ (d.o.f.)     &  \multicolumn{2}{c}{0.96 (431)} &  \multicolumn{2}{c}{1.04 (485)} &  \multicolumn{2}{c}{1.17 (511)} \\
\hline
        \end{tabular}
	\begin{tablenotes}
        \item $^{a}$ Here $T_{\rm p}$, $\tau_{\rm p}$ and $T_0$ are the plasma temperature, plasma optical depth and temperature of the seed photons for the {\sc comptt} model, respectively. Fluxes are given  in the 0.5--100 keV energy range.  Parameters of the iron line, modelled with the Gaussian, were fixed at the energy of $E_{\rm Fe}=6.4$~keV and width of 0.1 keV.
        \end{tablenotes}
\end{center}
\end{table*}

In all three {\it NuSTAR} observations we detected strong pulsations from \a05\
with the pulsed fraction decreasing from $\sim50$ to $\sim20$ per cent
following the decrease of the flux.
The pulse profile also evolves with luminosity becoming more single peaked at lower flux.
In either case we did not detect significant phase lags between relatively soft (3-13
keV) and hard (13-79 keV) pulse profiles expected in the case of different
geometrical structures responsible for emission in these energy bands.

In contrast to the timing properties, the source spectrum  strongly depends on the accretion rate. 
Spectra in the two brighter observations were already discussed by \cite{2017A&A...608A.105B}. Particularly, they showed that the spectrum of the bright state at luminosity $2\times10^{36}$ \lum\  can be fitted with a single either empirical or physically motivated continuum model. However, when the luminosity dropped down to $0.6\times10^{36}$ \lum, strong residuals started to appear around 30~keV, depending on the assumed continuum shape. The authors claimed that only {\sc plcut} model was able to describe the data satisfactorily. 
 However, an additional broad Gaussian emission component centered at $\sim26$~keV was required. As we discuss below,
this is not the only way to describe the broadband continuum, however, it illustrates nicely that some hard excess emission
appears already at intermediate fluxes.

\begin{figure}
\includegraphics[width=0.97\columnwidth, bb=30 270 555 710]{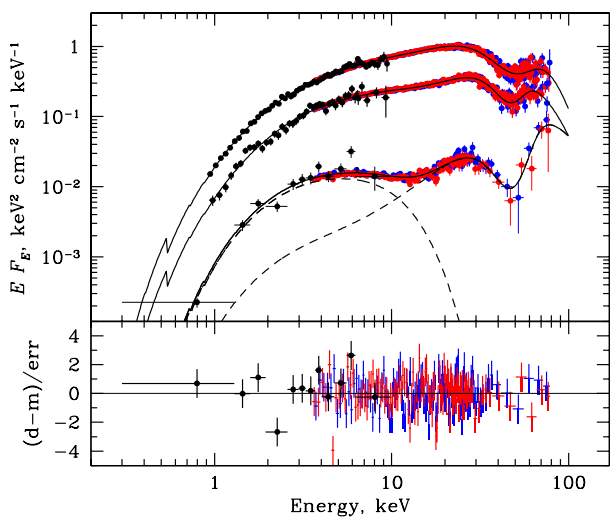}
\caption{$EF_{\rm E}$ spectra of \a05\ in different states with luminosity varying by almost two orders of magnitude, from $2\times10^{36}$ and $0.6\times10^{36}$ \lum\ for the  top two spectra down to $\sim7\times10^{34}$ \lum\ for the bottom one. 
The black, red and blue points correspond to the {\it Swift}/XRT and {\it NuSTAR} FPMA and FPMB data, respectively. 
The solid lines represent the best-fitting models listed in Table~\ref{tab:spe}. Dashed lines represent two {\sc comptt} components constituting the emission continuum. 
The corresponding residuals for the lowest luminosity state of \a05\ are presented at the lower panel. 
}
 \label{fig:specs}
\end{figure}

In our {\it NuSTAR} observation the source luminosity dropped further by an order of magnitude to $\sim7\times10^{34}$ \lum\ and its spectrum changed dramatically (see Fig.~\ref{fig:specs}).
Following \cite{2019MNRAS.483L.144T}, and to facilitate comparison of the spectra in all three observations and between
the sources, we fitted all three spectra with a two-component continuum consisting of two Comptonization components ({\sc comptt}) modified with the interstellar absorption ({\sc phabs}), fluorescent iron line ({\sc gauss}) and cyclotron line with Gaussian profile ({\sc gabs}). The final model is {\sc phabs$\times$(gauss+comtt+comptt)$\times$gabs} in {\sc xspec} with best-fitting parameters presented in Table~\ref{tab:spe}. The temperature of the seed photons was tied for the low- and  the high-energy components. The position and the width of the iron emission line was fixed at 6.4 keV and 0.1 keV, respectively.

\begin{figure}
\centering
\includegraphics[width=0.97\columnwidth, bb=30 305 550 710]{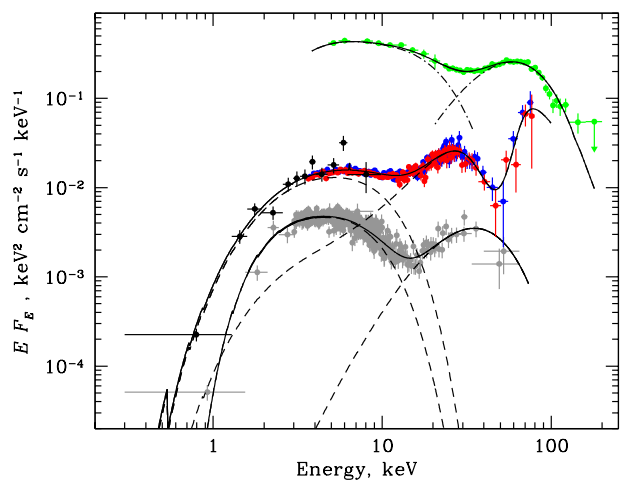}
\caption{The spectra of X~Persei  (green points; \citealt{2012A&A...540L...1D}), \gx\ (grey points; \citealt{2019MNRAS.483L.144T}) and \a05\ (this work) observed at low accretion rates,
along with the best-fitting models consisting of two Comptonization components (black solid lines).
 Separate model components ({\sc comptt} in {\sc xspec}) are shown with dashed lines.
 }
 \label{fig:persspec}
\end{figure}

\section{Discussion}
\label{sec:discus}

\subsection{Origin of the high-energy component}

The broadband spectra of the three X-ray pulsars observed so far in the low
state presented in Fig.~\ref{fig:persspec} show striking similarity. Particularly, in all three
cases the spectrum can be described as a blend of two broad components peaking
around 5--7 keV and 30--50 keV, respectively. We note that while the single
component cutoff power law modified by a broad absorption feature at
$\sim30$\,keV interpreted as a cyclotron line had been invoked previously to
describe the spectrum of X~Persei \citep{2001ApJ...552..738C,2012MNRAS.423.1978L}, this approach
actually appears to be unjustified both for X~Persei \citep{2012A&A...540L...1D} and \gx\ 
\citep{2019MNRAS.483L.144T}. It is even more true for \a05 where the
cyclotron line is also observed on top of the hard component, which completely
rules out such interpretation for this source. Given the overall similarity of spectra of the
three sources, we conclude thus that the hard component is not artificially
appearing because of the presence of a broad cyclotron line at intermediate energies,
but rather represents a physically independent component.

The question is then, what is the origin of the two components. From
observational point of view, both appear to be consistent with Comptonization
by relatively cool ($kT\sim2.5$\,keV) and hot ($kT\sim15$\,keV) electrons.
While the soft component is readily associated with thermal Comptonization
within the accretion-heated hotspot, the origin of the hard component is less
clear. As discussed by \citet{2019MNRAS.483L.144T}, bulk Comptonization of soft seed photons in
the accretion channel suggested previously for X~Persei \citep{2012A&A...540L...1D} is
unlikely to be effective because of the extremely low optical depths in the
accretion channel at the observed accretion rates. We note that this also likely
applies to any kind of Comptonization of soft seed photons unless their source
originates deep in the atmosphere of the neutron star.

Another interpretation discussed by \cite{1998ApJ...509..897D} and
\cite{2001ApJ...552..738C} for X~Persei is based on idea first suggested by
\cite{1995ApJ...438L..99N}. In this scenario fast ions of the accretion flow
decelerate in the atmosphere mostly by collisional excitation of
atmospheric electrons to the upper Landau levels. These then quickly recombine
emitting photons at the cyclotron energy which propagate through the atmosphere
losing the energy by the electron recoils, and ultimately produce a soft
thermal continuum plus a hard non-thermal line just below the cyclotron
energy. 
\citet{2001ApJ...552..738C} dismissed this idea based on the fact
that observed spectrum of X~Persei is well described with a broad component rather
than a comparatively narrow line with a sharp cutoff above the cyclotron
energy predicted by \citet{1995ApJ...438L..99N}. \citet{2012A&A...540L...1D}
also noted that luminosity of the hard component in X~Persei, which constitutes
$\sim30$ per cent of the total luminosity is substantially higher than that predicted by
\citet{1995ApJ...438L..99N}, thus also rejecting this scenario based on
energetic arguments. 
We note, however, that \citet{1995ApJ...438L..99N}
considered cold atmosphere with the temperature negligible compared to the cyclotron
energy. This simplifies the calculations but also amplifies energy losses of
high energy photons as they diffuse through the atmosphere, and thus diminish
the contribution of the hard component. The qualitative picture discussed below is essentially an extension of the scenario by \cite{1995ApJ...438L..99N} with the assumption of the low-temperature atmosphere dropped.

Indeed, to keep the atmosphere cool, the energy deposited by the collisional
excitation of electrons into Landau levels must still escape from the emission
region, predominantly through thermal free-free emission. However, based on the
analogy with the accretion-heated magnetized and non-magnetized NS atmospheres \citep[see,
e.g.][and discussion therein]{2000ApJ...537..387Z, 2001A&A...377..955D, 2018A&A...619A.114S,2019MNRAS.483..599G}, 
the density in the upper atmosphere layers is likely insufficient to make this
process effective. The upper atmosphere layers are thus expected to be
overheated, and thermal electron energies may actually be comparable with the
cyclotron energy.
In this case the energy losses of resonant photons through
recoil will be negligible, and we can anticipate a larger fraction of flux to
escape around the cyclotron energy in a broad component with the  
width defined by the temperature of the atmosphere. We note
  increased temperature of the hard component at low luminosity (see Table~\ref{tab:spe}), which is likely associated with the increased
  temperature of the upper atmosphere layers (Mushtukov et al., in preparation).
A narrow absorbtion
feature on top of this component may still be produced in the accretion flow just above the NS surface.
Indeed, the Thomson optical thickness of the accretion channel at
$L\sim 7\times 10^{34}\,{\rm erg\,s^{-1}}$ is $\tau_{\rm T}\sim 10^{-2}$, but
the optical thickness to the resonant scattering is $\tau\gtrsim 10^2$,
which is more than enough to explain the observed absorption feature around 47 keV.

It is important to emphasize that while the qualitative scenario outlined above
has to be confirmed with detailed numerical computations, it appears that
collisional excitation of electrons in strong magnetic field is the only way to
deposit the power of the accretion flow to higher photon energies where a significant
fraction of the observed flux emerges. The results presented above show that the 
presence of a hard component appears quite common in low-luminosity XRPs, 
therefore observing additional sources with a range of magnetic fields
becomes now feasible.

\begin{figure}
\centering
\includegraphics[width=0.95\columnwidth, bb=40 265 550 685]{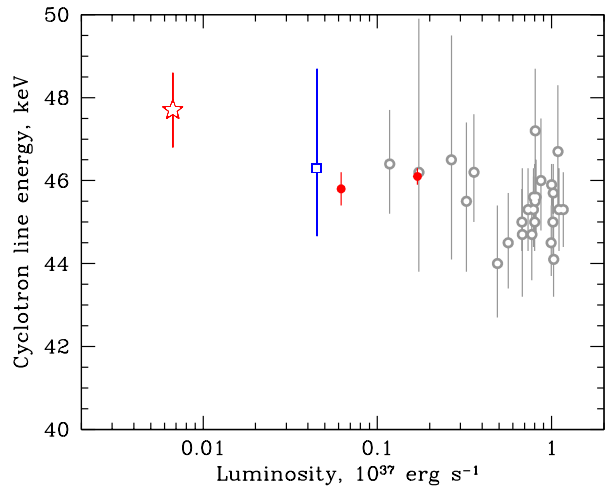}
\caption{The dependence of the cyclotron line energy in \a05\ on the bolometric (0.5--100 keV) luminosity.
The red star represents measurement in the lowest luminosity state, while the two filled red circles correspond to the brighter {\it NuSTAR} observations (see Table~\ref{tab:spe}). The blue open square and grey open circles show measurements taken from \citet{2006ApJ...648L.139T} and \citet{2007A&A...465L..21C}, respectively. 
 }\label{fig:ecyc}
\end{figure}

\subsection{Cyclotron line luminosity dependence}

Depending on the local mass accretion rate onto the magnetic poles of the NS, all XRPs can be roughly divided in two groups with different response of the cyclotron energy on the changing luminosity. In bright sources with fluxes high enough to stop accreting material above the NS surface, a negative correlation of the cyclotron energy with luminosity is expected \citep{2012AA...544A.123B,2013ApJ...777..115P} and observed in at least one bright transient XRP V~0332+53 \citep{2006MNRAS.371...19T}. In the opposite case, when luminosity of the source is below the critical one \citep[$L_{\rm crit}\sim10^{37}$ \lum; see ][]{1976MNRAS.175..395B,2015MNRAS.447.1847M} a positive correlation is observed in several sources \citep[][for more details, see a review by \citealt{2019A&A...622A..61S}]{2007A&A...465L..25S,2015MNRAS.454.2714M}. Transition between these two regimes in V~0332+53 was recently discovered by \cite{2017MNRAS.466.2143D}.

\a05\ has a relatively low luminosity rarely exceeding $\sim10^{37}$~\lum\ even in the maximum of giant outbursts. Therefore, the source belongs to the sub-critical group of XRPs where positive correlation can be expected.  During the last decade \a05\ was studied in great details with different instruments in order to detect such a correlation. However, in the pulse-averaged data it was not found in very broad range of mass accretion rates \citep[see, e.g.][]{2007A&A...465L..21C,2013ApJ...764L..23C,2017A&A...608A.105B}. 

Our {\it NuSTAR} observation allowed us to measure the cyclotron line energy in \a05\ with high accuracy at luminosity almost one order of magnitude below the lowest available in the literature. The results shown in Fig.~\ref{fig:ecyc} confirm the absence of the positive correlation of the cyclotron energy on the mass accretion rate in the source. Instead of positive correlation, some hint to the negative correlation of the cyclotron energy on the source luminosity is seen at the lowest luminosity. However, it has a very low significance, especially taking into account substantial change in the shape of the underlying continuum which can affect the deduced line energy. We thus confirm earlier findings and conclude that there is no strong evidence for correlation or anti-correlation of the line energy with luminosity in this source.

\section{Conclusion}

In the work we presented results of the broadband spectral analysis of emission from the transient XRP \a05, obtained for the first time with the {\it NuSTAR} observatory at the very low luminosity $\sim7\times10^{34}$~\lum. The data revealed that the spectrum can be described as a blend of two broad components peaking around 5--7 keV and 30--50 keV and thus differs dramatically from the cutoff power-law spectrum observed at higher fluxes.  
We explain the appearance of the high-energy component by recombination of electrons collisionally excited to the upper Landau levels in the heated  layers of the neutron star atmosphere. We were also able to accurately measure the cyclotron line energy $E_{\rm cyc}=47.7\pm0.8$~keV at a low luminosity never achieved before by the broad-band instruments. This allowed us to firmly exclude a positive correlation of the cyclotron energy with the accretion rate in this source.

\section*{Acknowledgements}
This work was supported by the Russian Science Foundation grant 19-12-00423.
We also acknowledge the support from the Academy of Finland travel grants 317552,  324550 (SST)  and  322779 (JP), 
the V\"ais\"al\"a Foundation (SST),  
the  Netherlands Organization for Scientific Research Veni Fellowship (AAM), the DFG grant WE 1312/51-1 (VFS) and the German Academic Exchange Service (DAAD) travel grant 57405000 (VFS). 
We grateful to the {\it NuSTAR} team for approving the DDT observation of \a05.


\bibliographystyle{mnras}
\bibliography{allbib}

\bsp    
\label{lastpage}
\end{document}